\documentclass{ws-procs9x6}
\begin{document}
\title{Photon and nucleon induced production of $\Theta^+$}
\author{Seung-Il Nam$^{1,2}$, Atsushi Hosaka$^1$ and Hyun-Chul
Kim$^{2}$}
\address{1. Research Center for Nuclear Physics (RCNP), \\Osaka
University, Ibaraki, Osaka
567-0047, Japan\\sinam@rcnp.osaka-u.ac.jp\,\,\,\, 
hosaka@rcnp.osaka-u.ac.jp\\and \\2. Nuclear physics \& Radiation technology
Institute (NuRI), 
\\Pusan University, Keum-Jung Gu, Busan 609-735, Korea \\hchkim@pusan.ac.kr} 
\maketitle
\abstracts{We investigate $\Theta^+$ production via photon
  and nucleon induced reactions.  We observe that the positive parity
  $\Theta^+$ production provides about ten times larger total cross
  sections than those of the negative parity one in both photon and nucleon
  induced reactions due to $P$--wave enhancement of the $KN\Theta$
  vertex. We also 
  consider the model independent method in the nucleon induced
  reaction to determine the  
  parity of $\Theta^+$ and show clearly distinguishable signals for the
  two parities.} 
\section{Introduction}
After the observation of the evidence of $\Theta^+$ by LEPS
collaboration~\cite{Nakano} motivated by Diakonov 
{\it et al.}~\cite{Diakonov}, physics of exotic pentaquark baryon
state has been scrutinized by huge amount of research activities. In
the present work, we investigate 
$\Theta^+$ production via photon and nucleon induced
reactions using Born diagrams with a pseudoscalar $K$ and
vector $K^*$--exchange included. For the nucleon induced reaction, we
consider the model independent method to determine the parity of
$\Theta^+$ which has not been confirmed yet by experiments. In
calculations, we assume that $\Theta^+$ has the quantum numbers of
spin 1/2, isospin 0 and the decay width $\Gamma_{\Theta\to
KN}=15$ MeV is used to obtain $KN\Theta$ coupling
constant~\cite{Nakano,Diakonov}. We perform calculations for both
parities of $\Theta^+$.   
\section{Photon induced reactions: $\gamma N\to \bar{K}\Theta^+$}
\begin{figure}[tbh]
\begin{tabular}{cc}
\resizebox{5.3cm}{2.7cm}{\includegraphics{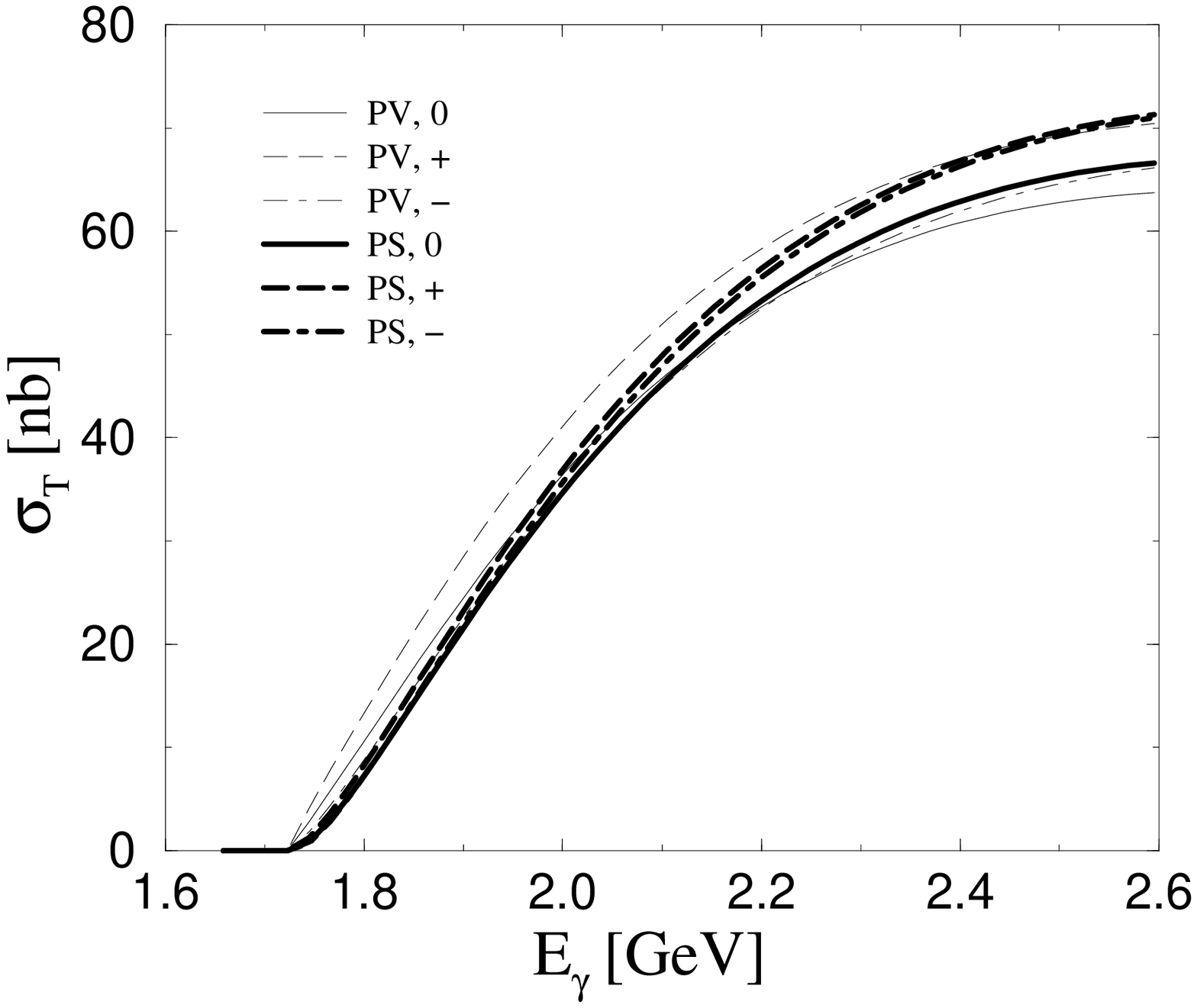}}
\resizebox{5.3cm}{2.7cm}{\includegraphics{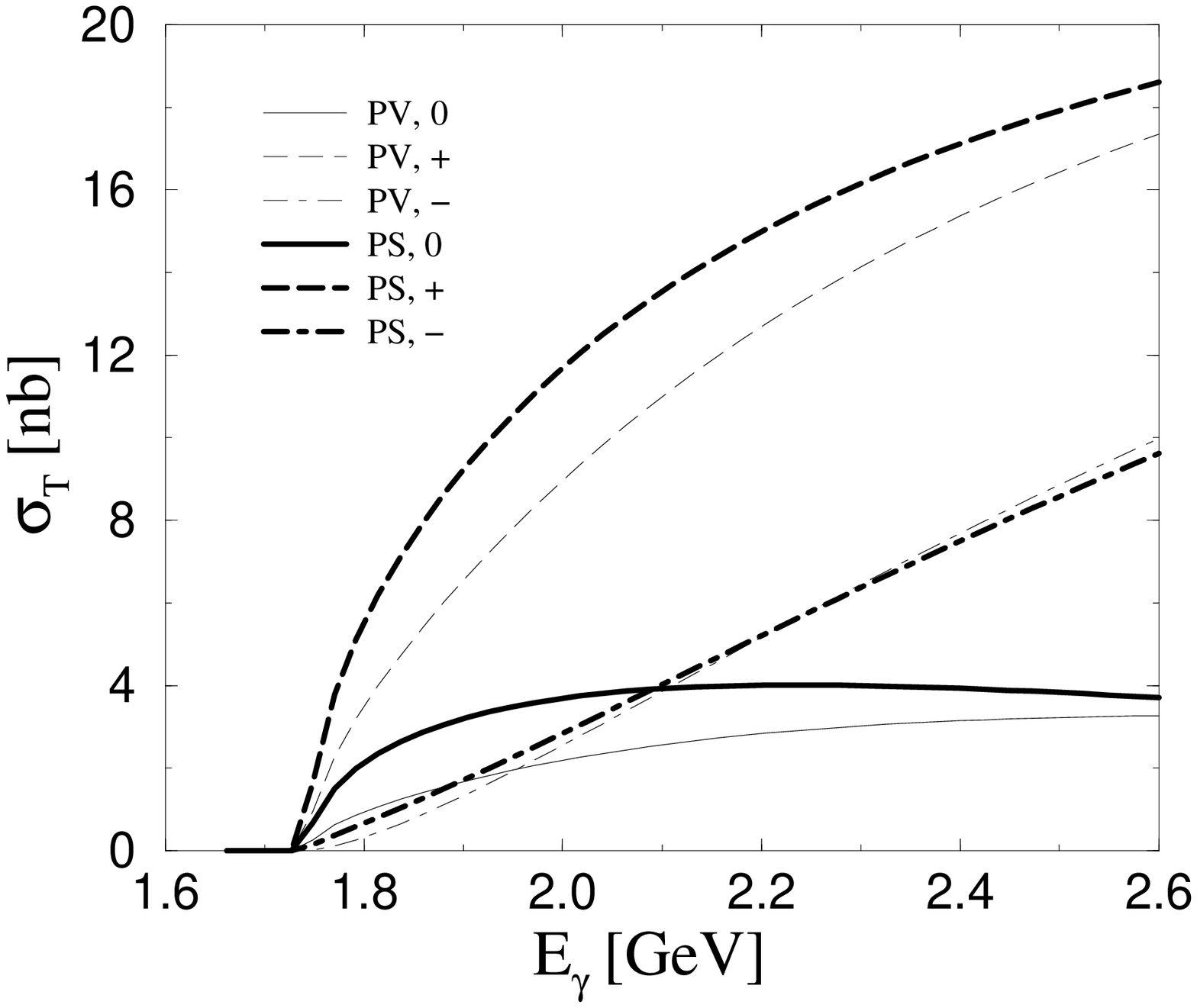}}
\end{tabular}
\caption{The total cross section of $\gamma n\to {K}^-\Theta^+$ (left)
  and   $\gamma p\to \bar{K}^0\Theta^+$ (right) for
  the positive parity. } 
\label{fig1}
\end{figure}

In this section, we study the total cross sections of $\gamma N\to
\bar{K}\Theta^+$ reactions. Results are given in
Fig~\ref{fig1}. Two models are employed for the $KN\Theta$ coupling
schemes. One is the  
pseudo-scalar (PS, thick lines) and the other is pseudo-vector (PV,
thin lines) to investigate theoretical ambiguity. As for the anomalous
magnetic moment of 
$\Theta^+$, $\kappa_{\Theta}$, we employ $-0.8$ considering several model
calculations~\cite{Nam2,Kim}.  We set the unknown $K^*N\Theta$ coupling
constant to be $|g_{KN\Theta}|/2$ with positive (dashed line)
and negative signs (dot-dashed line). In order to take into account
the baryon
structure, we employ a gauge invariant form factor
which suppresses $s$-- and $u$--channels~\cite{Nam2}. In 
Fig.\ref{fig1}  we plot total cross sections of the neutron (left) and
proton (right) 
targets only for the positive parity $\Theta^+$ since we
observe that the overall shapes and tendencies for the negative 
parity $\Theta^+$ are quite similar to the positive parity one.    
A major difference between them is that the total cross sections are
about ten times larger for the positive parity $\Theta^+$ than for the
negative parity one due to the $P$--wave coupling nature of the
$KN\Theta$ vertex. We also find that theoretical ambiguities due to the
PV and PS schemes, $\kappa_{\Theta}$ and $K^*$--exchange 
contribution become small for the 
neutron target where the $t$--channel $K$--exchange dominates, whereas
we find large model dependence for the proton case, where the
$K$--exchange does not appear.
\section{Nucleon induced reactions: $np\to Y\Theta^+$ and
$\vec{p}\vec{p}\to \Sigma^+\Theta^+$ } 
\begin{figure}[tbh]
\begin{tabular}{cc}
\resizebox{5.3cm}{2.7cm}{\includegraphics{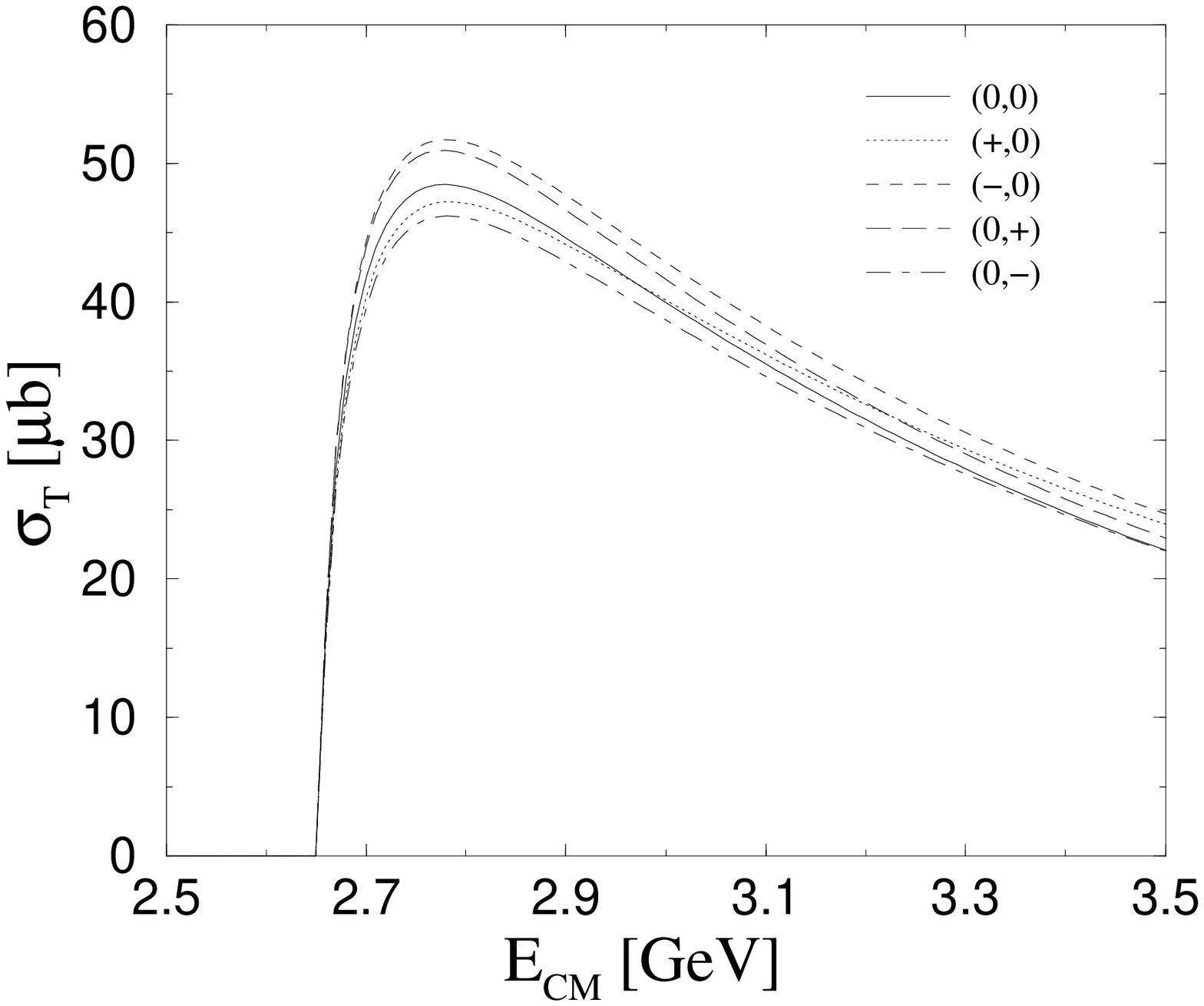}}
\resizebox{5.3cm}{2.7cm}{\includegraphics{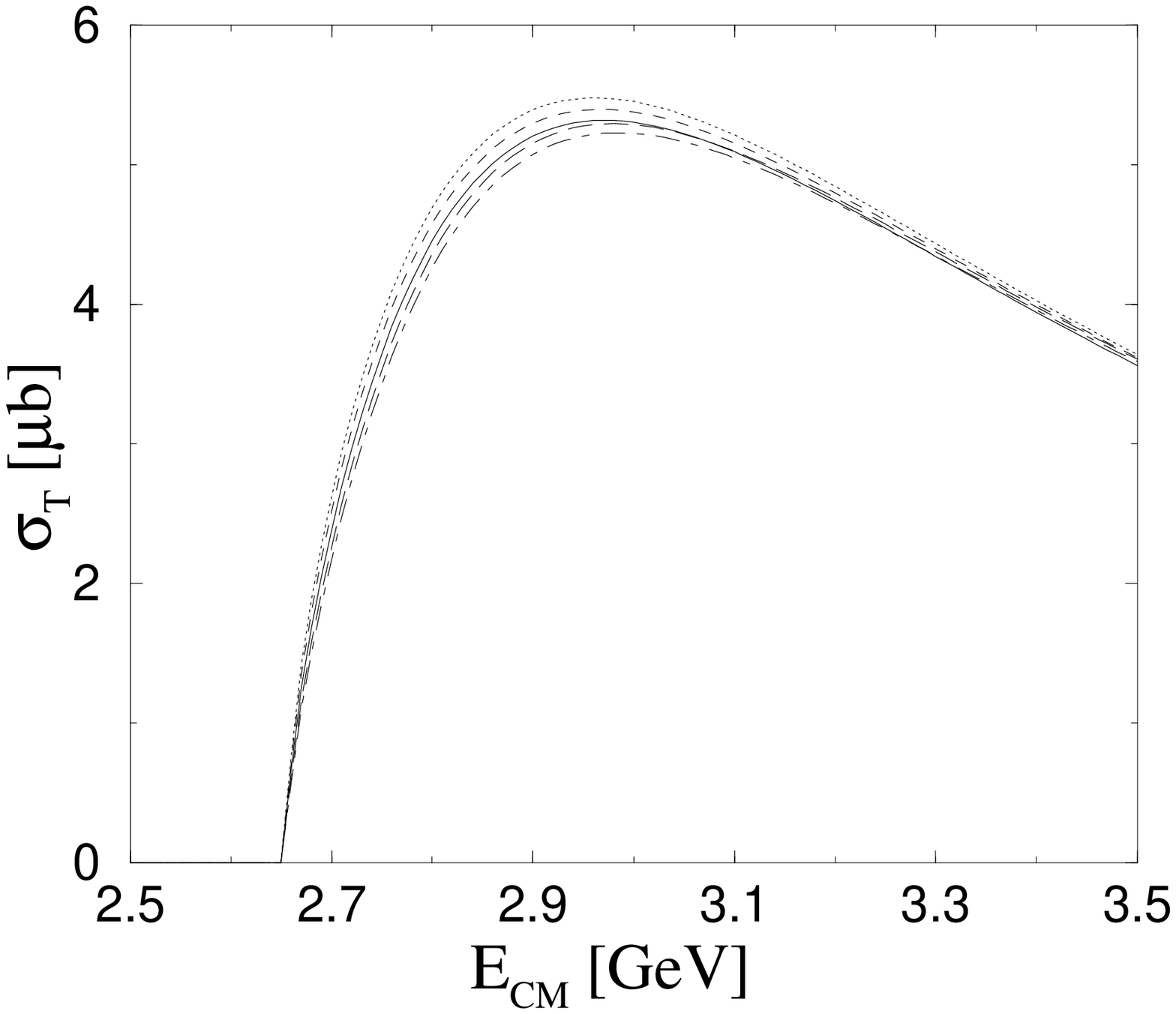}}
\end{tabular}
\caption{Total cross sections of $np\to \Lambda\Theta^+$ for the
  positive (left) and negative (right) parities of $\Theta^+$.} 
\label{fig3}
\end{figure}

In this section we investigate $NN$ scattering for the production of
$\Theta^+$. Here, we make use of 
the Nijmegen 
potential~\cite{stokes} for the $KNY$ coupling constants.  We
also take into account $K^*$--exchange 
contributions with vector and tensor $K^*N\Theta(Y)$
couplings~\cite{Nam5}. We consider only 
$Y=\Lambda$ since overall behaviors of $np\to \Sigma^0\Theta^+$ are
similar to $np\to \Lambda\Theta^+$ with differences in the order of
magnitudes of the total cross sections ($\sigma_{\Lambda}\sim 5\times
\sigma_{\Sigma^0}$).  We employ a monopole  
type form factor with a cutoff mass 1.0 GeV~\cite{Nam5}.  In
Fig.\ref{fig3} we plot the total cross sections for the reaction with
two different parities of $\Theta^+$. We observe that difference in 
the magnitudes of the total cross sections for the two parities
is similar to the photoproduction.  Furthermore, the
results are not very sensitive to the signs of vector and tensor
$K^*N\Theta$ coupling constants. The labels in parenthesis denote
(sgn($g^V_{K^*N\Theta}$),\,sgn$(g^T_{K^*N\Theta}$)). 
We note that if we consider 
initial state interaction, the order of magnitudes will be reduced by
about factor three~\cite{Nam5,Hanhart1}.
As suggested by Thomas {\it et al.}~\cite{Thomas}, taking into account
the Pauli principle and 
parity conservation,  $\vec{p}\vec{p}\to \Sigma^+\Theta^+$ provides a 
clear method for the determination of the parity of $\Theta^+$. Spin 0
initial state 
allows non-zero production rate near the threshold ($S$--wave) for the
positive parity $\Theta^+$, while spin 1 initial state does for the
negative parity one.  This
selection rule should not be affected by any model dependences.  We
confirm that at the threshold region ($\sim 2730$ MeV), the reaction
process is dominated by $S$-wave so that the selection rule is
applicable~\cite{Nam4}. 
\begin{figure}[tbh]
\begin{tabular}{cc}
\resizebox{5.3cm}{2.7cm}{\includegraphics{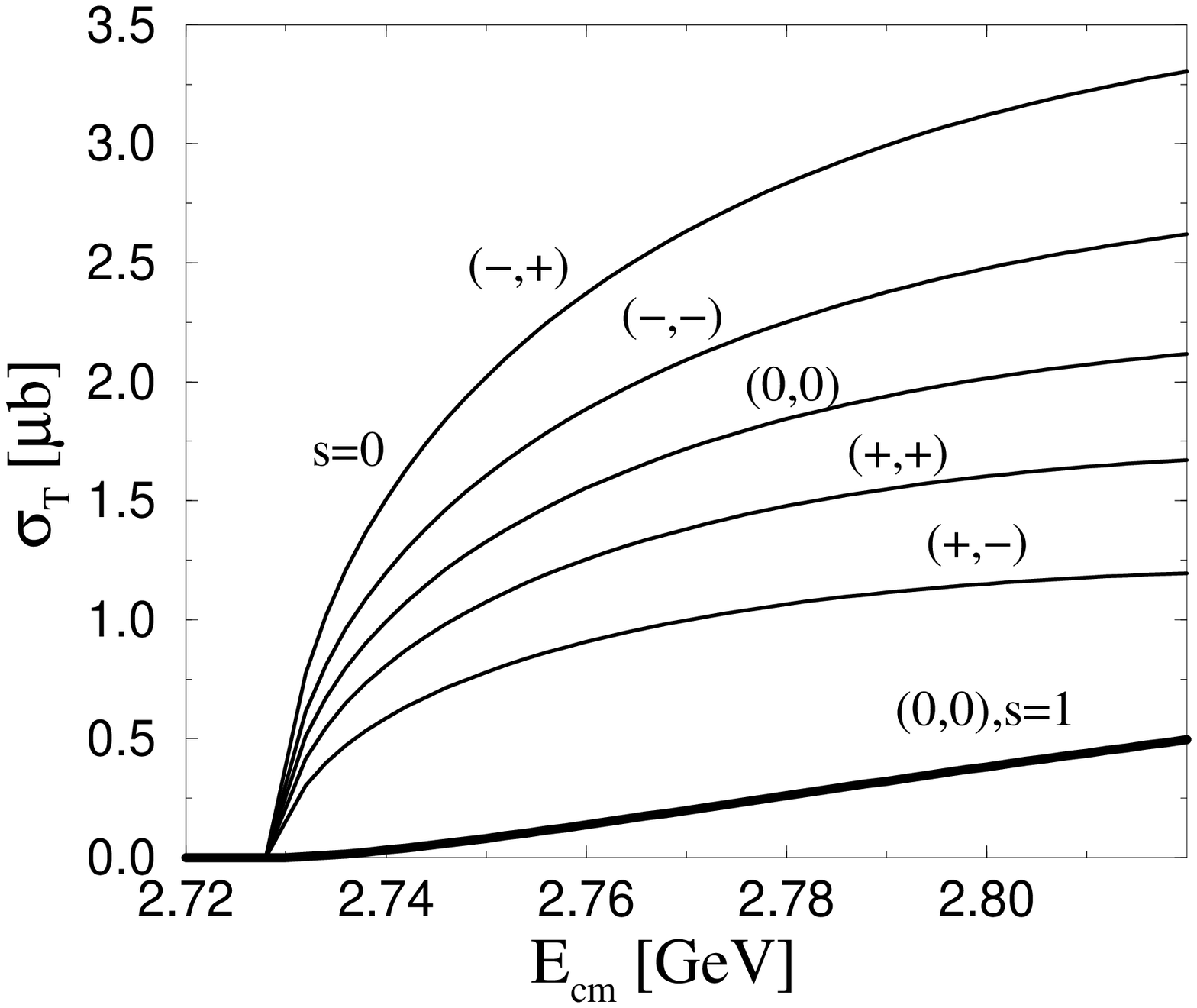}}
\resizebox{5.3cm}{2.7cm}{\includegraphics{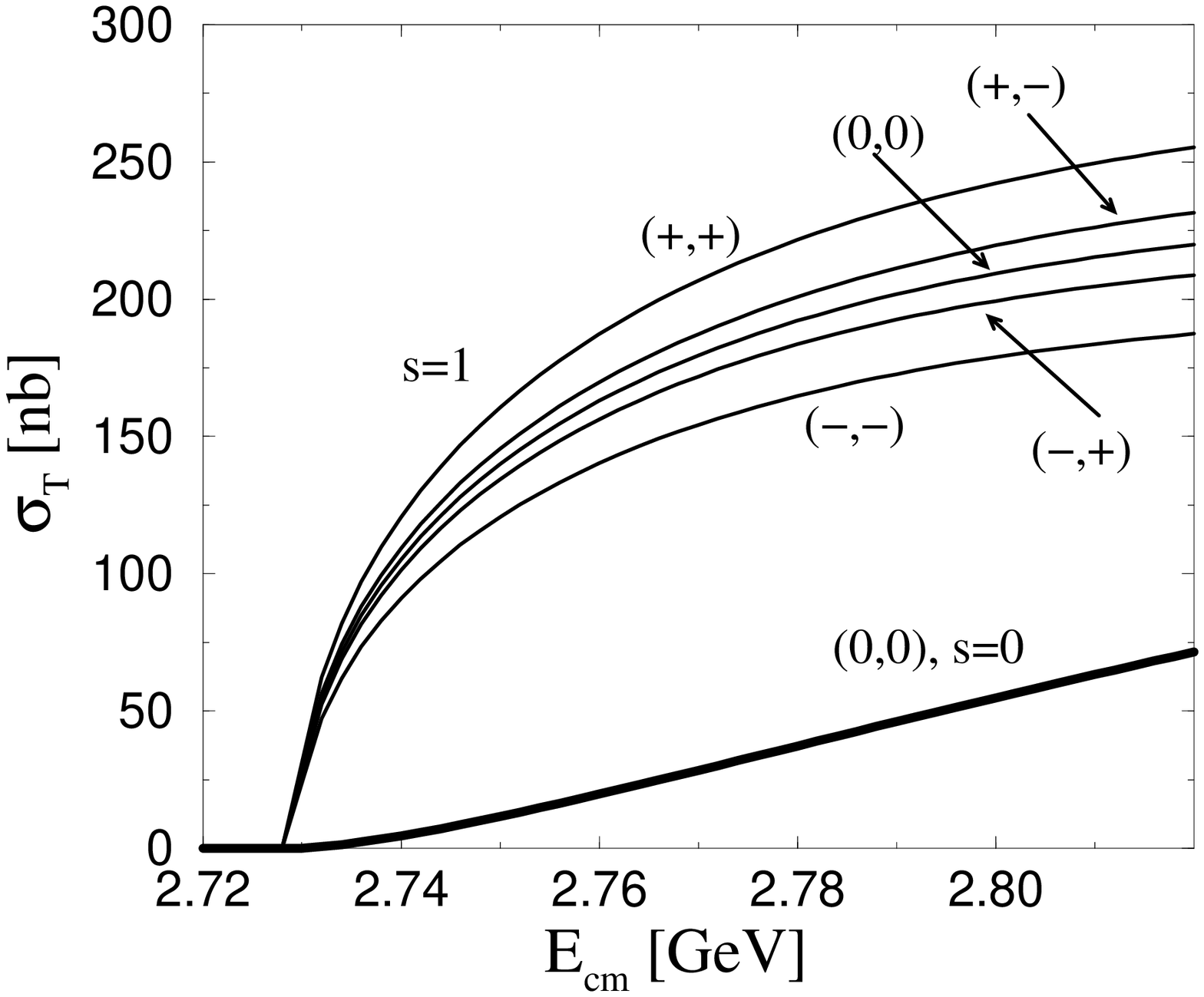}}
\end{tabular}
\caption{Total cross sections of $\vec{p}\vec{p}\to \Sigma^+\Theta^+$
  for the different spin states, spin 0 and spin 1 and for
the positive (left) and negative (right) parities of $\Theta^+$.}   
\label{fig4}
\end{figure}   
We observe clear evidences of the selection rule in
Fig.~\ref{fig4}.  $K^*$--exchange
contribution is not so sensitive to the various sign
combinations. The spin observable
$A_{XX}$, which was suggested by Hanhart {\it et al.}~\cite{Hanhart2}
is plotted 
in Fig.~\ref{fig5}. Up to about $100$ MeV above the threshold, the
results of the two
different parities of $\Theta^+$ show 
clear difference due to the selection rule~\cite{Nam4,Hanhart2}. 
\begin{figure}[tbh]
\begin{tabular}{cc}
\resizebox{5.3cm}{2.7cm}{\includegraphics{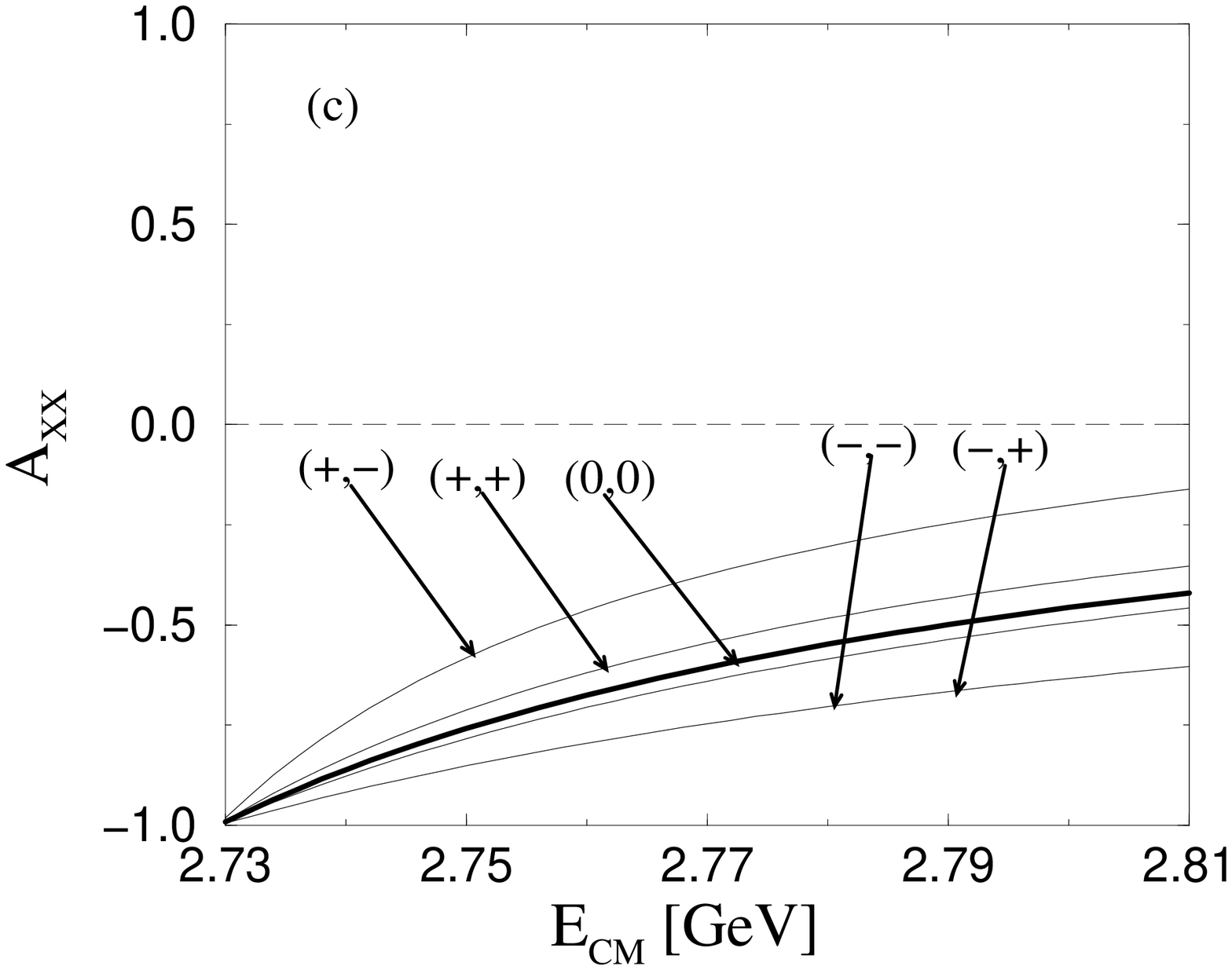}}
\resizebox{5.3cm}{2.7cm}{\includegraphics{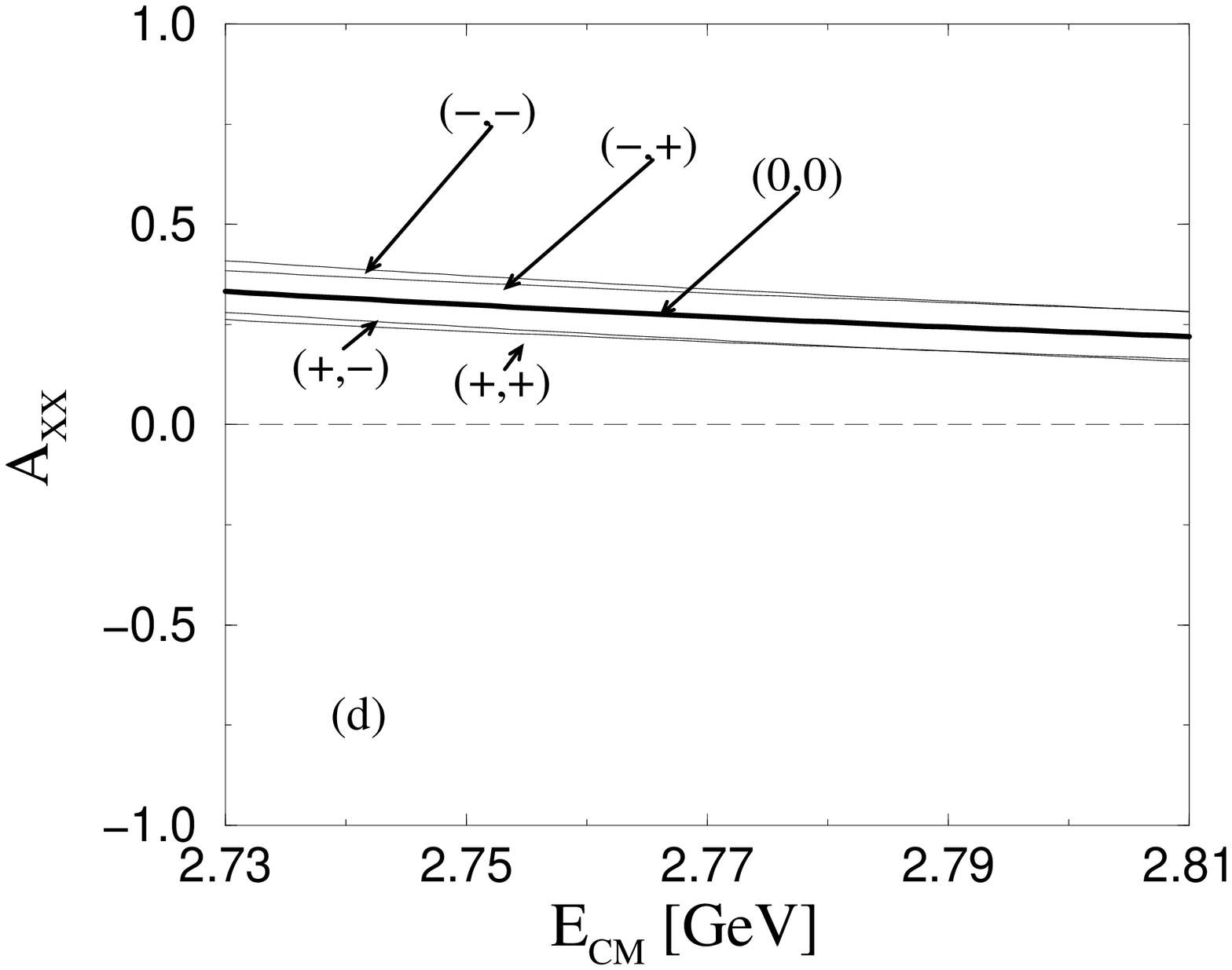}}
\end{tabular}
\caption{$A_{XX}$ for the positive (left) and negative (right) $\Theta^+$.} 
\label{fig5}
\end{figure}   
\section{Summary}
We have investigated the photon and nucleon induced reactions for the
$\Theta^+$ production in Born diagram calculations with
appropriate form factors and with some phenomenologically determined
coupling constants. Due to the different $KN\Theta$ vertex
structure, we observed about ten times larger total cross sections for
the positive parity $\Theta^+$. Though we still have  
theoretical ambiguities, this property is quite
universal for $\Theta^+$ production reactions. The model independent 
method to determine the parity of $\Theta^+$ via polarized $pp$
scattering seems quite promising. As announced by COSY-TOF
collaboration,  we hope to see more experimental results from
$\vec{p}\vec{p}$ scattering in the near future.

\end{document}